\documentclass[aps,prb,showpacs,twocolumn,superscriptaddress]{revtex4}

\usepackage{amsmath} 
\usepackage{amssymb} 
\usepackage{bm}
\usepackage{graphicx}

\begin{document}


\title{First-principles study of spontaneous polarization 
in multiferroic BiFeO$_3$}

\author{J.~B.~Neaton}
\altaffiliation{Present address: The Molecular Foundry, Materials Sciences
               Division, Lawrence Berkeley National Laboratory,
               Berkeley, CA 94720, USA}
\affiliation{Department of Physics and Astronomy, Rutgers University,
 Piscataway, New Jersey 08854-8019} 
\author{C.~Ederer}
\affiliation{Materials Research Laboratory and Materials Department,
 University of California, Santa Barbara, California, 93106} 
\author{U.~V.~Waghmare}
\affiliation{Nehru Center for Advanced Scientific Research, Jakkur,
 Bangalore 560 064, India}
\author{N.~A.~Spaldin}
\affiliation{Materials Research Laboratory and Materials Department,
 University of California, Santa Barbara, California, 93106} 
\author{K.~M.~Rabe}
\affiliation{Department of Physics and Astronomy, Rutgers University,
 Piscataway, New Jersey 08854-8019}


\begin{abstract}
The ground-state structural and electronic properties of ferroelectric
BiFeO$_3$ are calculated using density functional theory within the
local spin-density approximation and the LSDA+U method. The crystal
structure is computed to be rhombohedral with space group $R3c$, and
the electronic structure is found to be insulating and
antiferromagnetic, both in excellent agreement with available
experiments. A large ferroelectric polarization of
90-100~$\mu$C/cm$^2$ is predicted, consistent with the large atomic
displacements in the ferroelectric phase and with recent experimental
reports, but differing by an order of magnitude from early
experiments. One possible explanation is that the latter may have
suffered from large leakage currents. However both past and
contemporary measurements are shown to be consistent with the modern
theory of polarization, suggesting that the range of reported
polarizations may instead correspond to distinct switching paths in
structural space. Modern measurements on well-characterized bulk
samples are required to confirm this interpretation.
\end{abstract}

\date{\today}


\maketitle


\section{\label{sec:intro}Introduction}

There has been considerable recent interest in developing
multifunctional materials in which two or more useful properties are
combined in a single compound. Perhaps the most widely studied class
of multifunctional materials are the diluted magnetic semiconductors,
where interaction between magnetic and electronic degrees of freedom
allows both charge and spin to be manipulated by applied homogeneous
electric fields.\cite{ohno} These and other related materials having
spin-dependent electronic properties are presently being explored for
spintronic applications.\cite{spintronic} However another class of
materials, the so-called
multiferroics,\cite{smolenskii-chupis,schmid1} is also of growing
importance. Multiferroic materials have simultaneous ferromagnetic,
ferroelectric and/or ferroelastic ordering. Coupling between the
magnetic and ferroelectric order parameters can lead to {\it
magnetoelectric} effects, in which the magnetization can be tuned by
an applied electric field and vice versa. Relatively few multiferroics
have been identified,\cite{hill} and in those that are known, the
mechanism underlying their ferroelectricity is often
unconventional.\cite{YMnO3_Nature,Seshadri_Hill} The purpose of this
work is to understand the unusual ferroelectric behavior in
multiferroic BiFeO$_3$, which has recently emerged as an especially
promising magnetoelectric multiferroic material.\cite{wang}

Recently, large ferroelectric polarizations, exceeding those of
prototypical ferroelectrics BaTiO$_3$ and PbTiO$_3$, have been
reported in high quality thin films of
BiFeO$_3$.\cite{wang,RameshPC,yun} These sizeable polarizations are
consistent with the observed large atomic
distortions,\cite{michel,kubel} but apparently inconsistent with
earlier studies of {\it bulk} BiFeO$_3$,\cite{teague} a difference
whose origin is currently under debate. In addition, appreciable
magnetizations ($\sim$~1~$\mu_{\rm B}$/formula unit), increasing with
decreasing film thickness, have been reported,\cite{wang} accompanied
by substantial magnetoelectric coupling.

\begin{figure}
\includegraphics*[width=0.8\columnwidth]{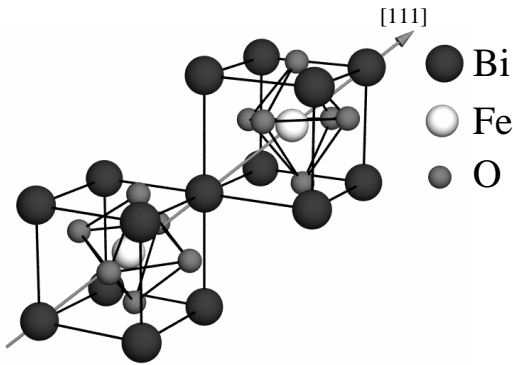}
\caption{Structure of $R3c$ BiFeO$_3$. Notice the position of the
oxygen octahedra relative to the Bi framework; in the ideal cubic
perovskite structure the oxygen ions would occupy the face-centered
sites.}
\label{fig:r3c-struc}
\end{figure}

Bulk BiFeO$_3$ has long been known to be ferroelectric \cite{teague}
with a Curie temperature of about 1100~K. The structure of the
ferroelectric phase, resolved experimentally using both X-ray and
neutron diffraction,\cite{michel,kubel} can be understood as
highly-distorted perovskite with rhombohedral symmetry and space group
$R3c$. The primitive unit cell contains two formula units (10 atoms) as
shown in Fig.~\ref{fig:r3c-struc}. The $R3c$ symmetry permits the
development of a spontaneous polarization along [111], and Bi, Fe, and
O are displaced relative to one another along this 3-fold axis.  The
largest relative displacements are those of Bi relative to O,
consistent with a stereochemically-active Bi lone
pair.\cite{Seshadri_Hill} The polar displacements (relative to cubic
perovskite) are noticeably extreme when compared with those in
non-lone-pair-active perovskite ferroelectrics such as BaTiO$_3$ or
KNbO$_3$, but are consistent with those observed in other Bi-based
perovskites.\cite{jorge} The observed counter-rotations of neighboring
octahedra about [111] are consistent with a tolerance factor somewhat
less than unity.  The crystal structure of the paraelectric phase has
not been conclusively determined.

Surprisingly, given the large atomic displacements relative to the
centrosymmetric cubic perovskite structure, and the high ferroelectric
Curie temperature, early measurements on bulk single
crystals\cite{teague} yielded rather small polarizations. Teague {\it
et al.} \cite{teague} initially reported a polarization along [111] of
just 6.1~$\mu$C/cm$^2$; although these authors state that their
hysteresis loops were not saturated, their measurements were supported
by more recent studies of both BiFeO$_3$/BaTiO$_3$ alloy
films\cite{ueda} and pure BiFeO$_3$ polycrystalline
films.\cite{palkar} The small values are in sharp contrast with recent
experiments on epitaxial thin film samples of BiFeO$_3$, which were
found to possess large polarizations. The first thin film
measurements\cite{wang} yielded values of 50-90~$\mu$C/cm$^2$ on (100)
aligned substrates, increasing to 100~$\mu$C/cm$^2$ for (111)
orientations.\cite{RameshPC} Following that, a whole variety of
different experimental values have been reported\cite{yun,ypwang,yun2}
(summarized in Table~\ref{tab:exp-values}), including a very recent
report \cite{yun} of giant ($> 150~\mu$C/cm$^2$) polarization which is
the highest value ever measured for a ferroelectric.

\begin{table}
\caption{Various measured values for the polarization in BiFeO$_3$, in
  chronological order with the oldest at the top.}
\label{tab:exp-values}
\begin{tabular}{c|c|l}
\hline\hline
& & \\
Ref. & $P$\ ($\mu$C/cm$^2$) & sample type\\ 
\hline
\hline 
\onlinecite{teague} & 6.1 & bulk single crystals \\
\hline
\onlinecite{ueda} & 2.5 & (Bi$_{0.7}$Ba$_{0.3}$)(Fe$_{0.7}$Ti$_{0.3}$)O$_3$
films (300~nm) \\ & & on Nb-doped SrTiO$_3$ \\
\hline
\onlinecite{palkar} & 2.2 & polycrystalline films (200~nm) \\
\hline
\onlinecite{wang} & 50 - 90 & thin films (400 - 100~nm) \\ & & 
on SrRuO$_3$/SrTiO$_3$ \\
\hline
\onlinecite{yun2} & 35.7 & polycrystalline films (350~nm) \\ \hline
\onlinecite{ypwang} & 8.9 & bulk ceramics \\ \hline
\onlinecite{yun} & 158 & polycrystalline films (300~nm) \\
\hline\hline
\end{tabular}
\end{table}

There are several plausible explanations for the spread of
experimental values. First, the original reports of small polarization
might have been limited by poor sample quality, with the large thin
film values representing the ``true'' polarization for $R3c$
BiFeO$_3$. A second possibility is that the small values could be
correct for the $R3c$ structure, with the large values being correct
for different structural modifications stabilized in the thin
films. And finally, a third possibility is that large and small values
can be explained within the modern theory of polarization, which
recognizes that polarization is in fact a lattice of values, rather
than a single vector.\cite{ksv,vks,resta} In this latter case, the
ferroelectric switching behavior in the different samples would have
to be substantially different. In this work, we use first-principles
density functional calculations to examine this issue carefully. We
find that the most natural value of the polarization is
90-100~$\mu$C/cm$^{2}$ along the [111] direction, consistent with
recent thin film measurements.\cite{wang,RameshPC} However both the
unexpectedly small early values, and the anomalously large recent
values, can be explained within the modern theory of polarization,
provided that a suitable switching path can be found.

The remainder of this paper is organized as follows. In
Section~\ref{sec:method} we describe the {\it ab initio} methods used
throughout this work. In Sections~\ref{subsec:structure} and
~\ref{subsec:electronic}, we report the computed ground state
structural and electronic properties of BiFeO$_3$ and show that they
are in good agreement with experiment. In
Section~\ref{subsec:polarization} the ferroelectric
polarization is calculated using the modern theory of polarization and compared with
a simple estimate. In section~\ref{subsec:connect} we discuss how
the dependence of the polarization on the switching path is a possible
explanation for the large spread of reported experimental values. The
intriguing magnetic properties are discussed in a separate
study.\cite{claude_inprep}

\section{\label{sec:method}Method}

To calculate the structure, polarization and Born effective charges of
BiFeO$_3$, we use density functional theory (DFT) within the local
spin-density approximation (LSDA) \cite{hk,ks} and the LSDA+U method
\cite{ldau} as implemented in the {\it Vienna ab initio Simulation
Package} (VASP).\cite{kresse,kresse2} All results were obtained using
the projector-augmented plane-wave (PAW) method \cite{paw,paw2} by
explicitly treating 15 valence electrons for Bi ($5d^{10}6s^26p^3$),
14 for Fe ($3p^63d^64s^2$), and 6 for oxygen ($2s^22p^4$). Our
scalar-relativistic calculations do not include spin-orbit
corrections. All structural relaxations are performed within the
LSDA. The ions are steadily relaxed toward equilibrium until the
Hellmann-Feynman forces are less than 10$^{-3}$~eV/{\AA}. Brillouin
zone integrations are performed with a Gaussian broadening
\cite{gaussian} of 0.1~eV during all relaxations. These calculations
are performed with a 3$\times$3$\times$3 Monkhorst-Pack {\bf k}-point
mesh \cite{Monkhorst/Pack:1976} centered at $\Gamma$ and a 500~eV
plane-wave cutoff, both of which result in good convergence of the
computed ground state properties.

Since it is well known that the LSDA often underestimates the size of
the band gap in systems with strongly localized $d$ orbitals, and even
predicts metallic behavior for materials that are known to be
insulators,\cite{Terakura_et_al:1984} we also calculate the electronic
structure (using the structural parameters obtained within the LSDA)
within the LSDA+U method.\cite{ldau} In the LSDA+U framework the
strong Coulomb repulsion between localized $d$ states is treated by
adding a Hubbard-like term to the effective potential, leading to an
improved description of correlation effects in transition metal
oxides.  The LSDA+U method requires two parameters, the Hubbard
parameter $U$ and the exchange interaction $J$. Since there is no
unique way of including a Hubbard term within the DFT-framework,
several different approaches exist which all give similar results. In
this work we use the approach described by Dudarev {\it et al.}
\cite{dudarev} where only an effective Hubbard parameter $U_\text{eff}
= U - J$ enters the Hamiltonian. The magnitude of $U_\text{eff}$ is
varied between 0~eV and 7~eV for the Fe $d$ states (the standard LSDA
result corresponds to $U_\text{eff}=0$~eV). These calculations are
performed with a $\Gamma$ centered 5$\times$5$\times$5 Monkhorst-Pack
{\bf k}-point mesh and a slightly lower plane-wave cutoff of
400~eV. The tetrahedron method \cite{Bloechl_tetrahedron:1994} is used
for Brillouin zone integrations.

Rock salt, or G-type, antiferromagnetic (AFM) order is assumed for all
calculations, as well as a homogeneous and collinear spin
arrangement. This assumption is well justified, since in practice
BiFeO$_3$ is observed to be nearly G-type AFM. However, experiments
also report a long wavelength spiral spin structure \cite{sosnowska}
and possibly a small out-of-plane canting due to weak
ferromagnetism.\cite{vorobev} Since the noncollinearity is quite
minimal, the simplification to a collinear magnetic structure is
acceptable. Detailed calculations of the effects of noncollinearity
and spin-orbit coupling on the magnetic properties of BiFeO$_3$ appear
separately.\cite{claude_inprep}

The electronic contribution to the polarization is calculated as a
Berry phase using the method first developed by King-Smith and
Vanderbilt\cite{ksv,vks} (see also Ref.~\onlinecite{resta}), the
so-called ``modern'' theory of polarization. In this approach, the
total polarization {\bf P} for a given crystalline geometry can be
calculated as the sum of ionic and electronic contributions. The ionic
contribution is obtained by summing the product of the position of
each ion in the unit cell (with a given choice of basis vectors) with
the nominal charge of its rigid core. The electronic contribution to
{\bf P} is determined by evaluating the phase of the product of
overlaps between cell-periodic Bloch functions along a densely-sampled
string of neighboring points in {\bf k}-space. Here we use 4
symmetrized strings consisting of 15 {\bf k} points to obtain the
electronic contribution to the polarization, which is calculated
separately for each spin channel; the total polarization is then the
sum of the two spin contributions and the ionic contribution.

\section{\label{sec:results}Results and Discussion}
\subsection{\label{subsec:structure} Structure}

In Table \ref{table:structure} we report structural parameters
obtained by relaxing the cell volume $\Omega$, rhombohedral angle
$\alpha$, and atomic positions within the $R3c$ space group, which are
all in good agreement with the experimental values of Kubel and
Schmid.\cite{kubel} The lattice constant is underestimated by roughly
3\%, a large but not atypical consequence of the LSDA.  The internal
atomic coordinates are reproduced very well.  Both the experimental
and the theoretical value are very close to 60$^\circ$ which would
correspond to perfectly cubic lattice vectors.  The calculated
rhombohedral angle is slightly larger than experiment; since it is
found to increase with decreasing volume, we attribute this to the
underestimated volume of the LSDA ground state.  Our calculations of
the total energy for other selected symmetries, including $P4mm$,
$R\bar{3}c$, and $R3m$, corroborate $R3c$ as the ground state.

\begin{table}[ht]
\caption{Calculated and measured structural parameters
of BiFeO$_3$ in space group $R3c$ (point group $C_{3v}$); the
Wyckoff positions 2a and 6b are referenced to the rhombohedral system and are
Bi($x,x,x$), Fe($x,x,x$), and O($x,y,z$). Also included are the
lattice constant $a_\text{rh}$ (rhombohedral unit cell), the
rhombohedral angle $\alpha$, and the unit cell volume $\Omega$.}
\begin{tabular}{lccc}
\hline\hline
\hspace{0.67in} & & \hspace{1.1in} & \hspace{1.1in}\\ & & LSDA &
 Exp.~(Ref.~\onlinecite{kubel})\\ \hline Bi (2a) & $x$ & 0 & 0 \\ Fe (2a) &
 $x$ & 0.231 & 0.221 \\ O (6b) & $x$ & 0.542 & 0.538 \\ & $y$ & 0.943
 & 0.933 \\ & $z$ & 0.408 & 0.395 \\ \hline $a_\text{rh}$ (\AA) & &
 5.46 & 5.63 \\ $\alpha$ $(^\circ)$ & & 60.36 & 59.35 \\ $\Omega$
 (\AA$^3$) & & 115.98 & 124.60 \\ \hline\hline
\end{tabular}
\label{table:structure}
\end{table}

In $R3c$, the Bi site is strongly distorted such that only 6
of the 12 oxygens surrounding Bi can still be considered nearest
neighbors; three co-planar oxygens lie above Bi along [111] at
2.30~{\AA}, and three sit below at 2.41~{\AA}. Likewise, the Fe site
is displaced relative to the center of its surrounding octahedron,
which is also distorted, with 3 oxygen neighbors at 1.92~{\AA} and 3
others at 2.07~{\AA}. The O-Fe-O bond angle in this system, which
would be an ideal 180$^\circ$ in a cubic perovskite structure, buckles
to a value of 165$^\circ$ in $R3c$. Indeed $R3c$ BiFeO$_3$ is
structurally similar to ferroelectric LiNbO$_3$, which also has $R3c$
symmetry, off-centering of the Li and Nb cations, and large rotations
of the oxygen octahedra resulting in a 6-fold coordinated site for the
large cation.\cite{cohen_lno,veithen}

The structural relationship between cubic perovskite and $R3c$ can be
understood with just two rather simple distortions from the cubic
geometry: (i) counter-rotations of adjacent oxygen octahedra {\it
about} [111], and (ii) relative ionic displacements {\it along}
[111]. Freezing in the counterrotations alone results in a nonpolar
insulator with $R\bar{3}c$ symmetry, a structural phase of BiFeO$_3$
that, by analogy to LiNbO$_3$, is a possible candidate for its
high-temperature paraelectric phase. Freezing in only the polar mode
produces a ferroelectric with $R3m$ symmetry.

\subsection{\label{subsec:electronic}Electronic and magnetic properties}

In this section, we present the electronic structure of BiFeO$_3$
computed in the ferroelectric $R3c$ structure. We also briefly discuss
results obtained for the cubic perovskite and centrosymmetric
$R\bar{3}c$ structures; both are possible paraelectric high-temperature phases.

\begin{figure}
\includegraphics*[width=0.8\columnwidth]{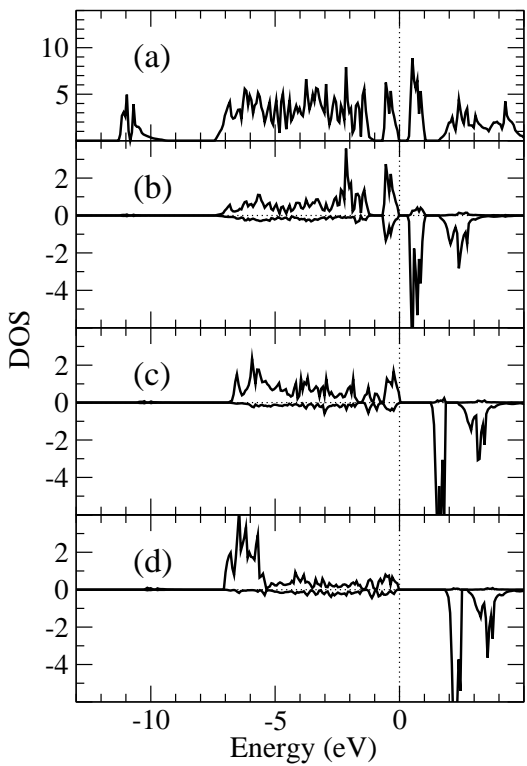}
\caption{Calculated densities of states (DOS) for $R3c$
BiFeO$_3$. Upper panels (a) and (b) show the total DOS for one spin
channel and the local Fe DOS for both spin channels (minority spin
states are shown as negative) calculated within the LSDA,
respectively.  Lower panels (c) and (d) show the local Fe DOS obtained
from the LSDA+U method with $U_\text{eff}$=2~eV and
$U_\text{eff}$=4~eV, respectively.  The zero is set to the valence
band maximum.}
\label{fig:r3c-dos}
\end{figure}

The single-particle density of states (DOS) for a single spin channel
and the DOS of the local Fe $d$ states for both spin channels,
calculated within the LSDA, are shown in Fig.~\ref{fig:r3c-dos}a 
and b for BiFeO$_3$ in the $R3c$ structure. Both spin channels exhibit
identical total DOS, as required for an AFM. The structure is
insulating, with a small gap of 0.4~eV in the LSDA calculation. This
gap is significantly enhanced after application of the LSDA+U method,
as seen from Fig.~\ref{fig:r3c-dos}c; for the small value
$U_\text{eff}$=2~eV, the gap is 1.3~eV; for $U_\text{eff}$=4~eV
(Fig.~\ref{fig:r3c-dos}d) it further increases to 1.9~eV. The narrow
bands around the Fermi energy arise predominantly from the Fe $3d$
states (with some oxygen hybridization) and are divided into $t_{2g}$
and $e_g$ manifolds, as expected from their octahedral
coordination. Lying below and hybridized with the Fe states is the
broad predominantly oxygen $2p$ valence band, which also contains a
significant amount of Bi $6p$ character. The lowest band shown, at
$\sim$~9.5~eV below the valence band maximum, is the Bi $6s$ band. The
large amount of occupied Bi $6p$ states is consistent with the picture
of the Bi lone pair as the driving force of the ferroelectric
distortion \cite{Seshadri_Hill,hill} in this class of materials.

\begin{figure}
\includegraphics*[width=0.9\columnwidth]{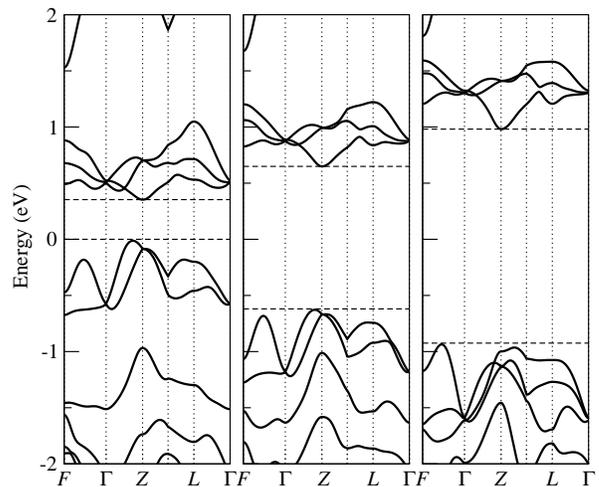}
\caption{Band structure of $R3c$ BiFeO$_3$ in the energy region of the
  gap for $U_\text{eff}$=0~eV (LSDA, left panel), $U_\text{eff}$=2~eV
  (middle panel), and $U_\text{eff}$=4~eV (right panel). The valence
  and conduction band edges are indicated by the dashed horizontal
  lines. The high-symmetry {\bf k}-points are labeled according to
  Ref.~\onlinecite{Bradley/Cracknell}.}
\label{fig:r3c-bands}
\end{figure}

Fig.~\ref{fig:r3c-bands} shows the dispersion of the bands in the
energy range around the gap. As can be seen in the figure, the band
gap is indirect for LSDA ($U_\text{eff}$=0~eV) and
$U_\text{eff}$=2~eV, with the bottom of the conduction band located at
the point $Z$ in the rhombohedral Brillouin zone and the top of the
valence band between $\Gamma$ and $Z$. For a higher value of
$U_\text{eff}$=4~eV, the gap remains indirect but the top of the
valence band shifts to a location between $\Gamma$ and $F$.
Photoemission spectroscopy would be of use to determine the
value of $U_\text{eff}$ best describing the electronic structure of
BiFeO$_3$.

The LSDA local magnetic moment at the Fe atoms (integrated within a
sphere of radius 1.4~{\AA}) is $\sim$3.3~$\mu_\text{B}$, in reasonable
agreement with the experimental value of 3.75~$\mu_B$
(Ref.~\onlinecite{sosnowska2}). It is reduced from the formal value of
5~$\mu_B$ for high spin Fe$^{3+}$ because of the finite bandwidth of
the 3$d$ states. Use of the LSDA+U method improves the agreement with
the experiment, enhancing the Fe magnetic moment to a value of
3.8~$\mu_\text{B}$ for $U_\text{eff}$=2~eV and 4.0~$\mu_\text{B}$ for
$U_\text{eff}$=4~eV.

\begin{figure}
\includegraphics*[width=0.8\columnwidth]{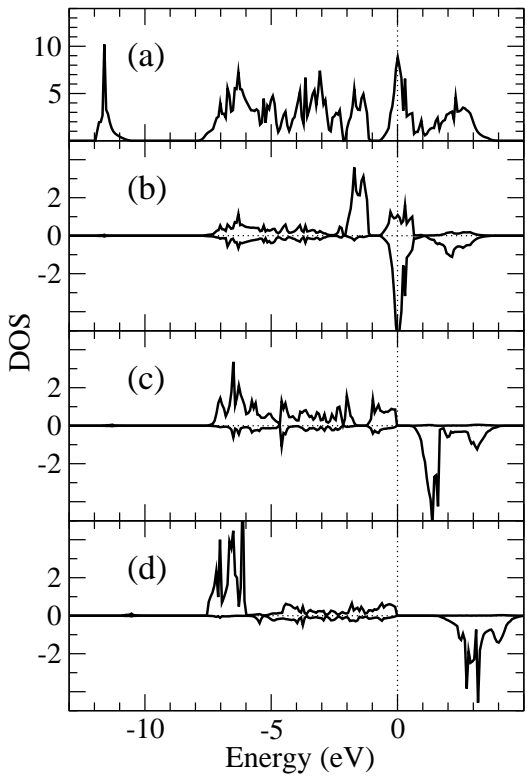}
\caption{Calculated densities of states (DOS) for cubic
BiFeO$_3$. Upper panels (a) and (b) show the total DOS for one spin
channel and the local Fe DOS for both spin channels respectively
(minority spin states are shown with a negative sign), both calculated
within the LSDA. Lower panels (c) and (d) show the local Fe DOS
obtained using the LSDA+U method, with $U_\text{eff}$ values of 2~eV
and 5~eV. The zero is set to the Fermi energy (a,b) or valence band
maximum (c,d).}
\label{fig:cubic-dos}
\end{figure}

For comparison and to discuss the origin of the ferroelectric
instability, we provide corresponding results for the high symmetry
cubic perovskite structure. A primitive lattice constant of 3.87~\AA\
is used for all calculations and corresponds to a volume per formula
unit equal to that of the bulk $R3c$ phase; the unit cell is doubled
along [111] to accommodate the opposite spin polarization of the two Fe
ions.  The single-particle total density of states (DOS), calculated
within the LSDA, is shown for one spin channel in
Fig.~\ref{fig:cubic-dos}a, along with the local density of states for
both spin channels on one of the Fe ions (Fig.~\ref{fig:cubic-dos}b).
Strikingly, BiFeO$_3$ is metallic in this structure within the LSDA,
with the Fermi energy cutting through the narrow band composed of
up-spin $e_g$ states and down-spin $t_{2g}$ states.

As discussed in Sec.~\ref{sec:method}, the LSDA often fails to
describe such narrow $d$ bands correctly. Fig.~\ref{fig:cubic-dos}c
and d show the site-projected local DOS for Fe obtained within the
LSDA+U method for two different values of $U_\text{eff}$. Including an
effective Coulomb repulsion parameter pushes the Fe majority $d$-bands
down and the Fe minority $d$-bands up in energy, leading to a gap of
$\sim$0.5~eV for a relatively small $U_\text{eff}$ of 2~eV. The
fully-gapped insulating phase is accompanied by a marked increase in
the local Fe magnetic moment from $\sim$1.9~$\mu_\text{B}$ to
$\sim$3.9~$\mu_\text{B}$ (indicating a low- to high-spin
transition). Typical values for $U_\text{eff}$ in other Fe compounds
range between $U_\text{eff}$=3~eV and
5~eV;\cite{Anisimov-Elfimov:1996,Anisimov-Gunnarsson:1991} the fact
that such a small $U_\text{eff}$ opens a gap supports the notion that
inadequate treatment of strong correlation in LSDA results in a
metallic state for cubic BiFeO$_3$.

Finally, we remark that the centrosymmetric $R\bar{3}c$ structure,
mentioned briefly in Section ~\ref{subsec:structure} as a possible
candidate paraelectric phase, is already insulating within the LSDA,
albeit with a small gap whose magnitude is found to be extremely
sensitive to input parameters.  As with the cubic perovskite phase, a
stable insulating solution can be obtained even for small values of
$U_\text{eff}$. We perform a full structural relaxation
imposing $R\bar{3}c$ symmetry within the LSDA, and the structural
parameters resulting from this relaxation are summarized in
Table~\ref{table:rbar3c}. Briefly, compared with $R3c$, this
arrangement is found to possess a smaller volume and a larger
rhombohedral angle.  The environment of the A site is drastically
changed relative to cubic perovskite: each Bi site has 3 oxygen
neighbors at 2.30~\AA, 6 at 2.71~\AA, and 3 more at 3.21~\AA. However
the local octahedral environment of the B site is preserved: each Fe
site is octahedrally coordinated with 6 O neighbors at a distance of
1.97~\AA.

\begin{table}[ht]
\caption{Calculated structural parameters (lattice constant
  $a_\text{rh}$, rhomobohedral angle $\alpha$, and volume $\Omega$) of
  BiFeO$_3$ in the space group $R\bar{3}c$ (point group D$_{3d}$); the
  Wyckoff positions, referenced to the rhombohedral system, are (2b)
  Bi(${1\over 2},{1\over 2},{1\over 2}$), (2a) Fe($0,0,0$), and (6e)
  O($x,{1\over 2}-x,{1\over 4}$).}
\begin{tabular}{lcc}
\hline
\hline
\hspace{0.67in}     & \hspace{1.1in}\\
                    & LSDA          \\
\hline
$x$                 & 0.417         \\
\hline
$a_\text{rh}$ (\AA) & 5.35          \\
$\alpha$ $(^\circ)$ & 61.93         \\
$\Omega$ (\AA$^3$)  & 113.12        \\
\hline
\hline
\end{tabular}
\label{table:rbar3c}
\end{table}

\subsection{Spontaneous polarization}
\label{subsec:polarization}

\subsubsection{Estimate using Born effective charges}

As discussed above, recent reports of ferroelectric polarizations in
high quality BiFeO$_3$ thin films\cite{wang,yun} exceed previous
measurements on bulk samples\cite{teague} by an order of
magnitude. The polarizations measured in thin film samples are
consistent with the observed large atomic
distortions,\cite{michel,kubel} but apparently inconsistent with an
earlier study of {\it bulk} BiFeO$_3$,\cite{teague} which reported a
modest polarization along [111] of just 6.1~$\mu$C/cm$^2$.

In an effort to shed light on this issue, we examine the spontaneous
polarization, ${\bf P}$, of BiFeO$_3$ in its $R3c$ ground state from
first principles. Experimentally, ${\bf P}$ corresponds to half the
polarization change as the applied field is swept through zero from
positive to negative. For prototypical perovskite ferroelectrics, it
has been standard to estimate this value by simply summing the product
of atomic displacements (from a centrosymmetric reference structure)
and their corresponding Born effective charges (BECs). This estimate
corresponds to computing the Cartesian components of the polarization
$\Delta P_\alpha$ to linear order in the atomic displacements, i.e.
\begin{equation}
\Delta P_{\alpha} \cong \sum_{j\beta} \frac{\partial
P_{\alpha}}{\partial u_{j\beta}} (u_{j\beta} -
u_{0j\beta})=\frac{e}{\Omega}\sum_{j\beta} Z^*_{j\alpha\beta} \Delta
u_{j\beta},
\label{eqn:polchange}
\end{equation}
where $\Delta u_{j\beta}$ is the displacement of ion $j$ in Cartesian
direction $\beta$, $Z^*_{j\alpha\beta}$ is its Born effective charge
tensor, and $\Omega$ is the unit cell volume. The zero
subscript generally refers to an insulating centrosymmetric reference
structure (in this case, either cubic perovskite or $R\bar{3}c$).
The spontaneous polarization is obtained from Eq.~(\ref{eqn:polchange}) 
by considering a specific structural (or {\it switching}) pathway parametrized by the change
in atomic displacements connecting a centrosymmetric reference structure 
and $R3c$.

\begin{table}
\caption{Born effective charges (BECs) for displacements along [111]
  for BiFeO$_3$ in the cubic perovskite, $R\bar{3}c$, and $R3c$
  structures. LSDA+U results are obtained using $U_\text{eff} =
  2$~eV. Since cubic perovskite is metallic within LSDA and the
  $R\bar{3}c$ structure is nearly so, LSDA values are given only for
  the $R3c$ structure. $\Delta P$, calculated using
  Eq.~(\ref{eqn:polchange}) along a path from the cubic structure to
  $R3c$, is given for each set of BECs.}
\label{table:BECs}
\begin{tabular}{lc|ccc|ccc}
\hline\hline 
& & \hspace*{.9cm} & \hspace*{.9cm} & \hspace*{.9cm} & \hspace*{.9cm}
& \hspace*{.9cm} & \hspace*{.9cm} \\
& & \multicolumn{3}{c|}{LSDA} & \multicolumn{3}{c}{LSDA+U} \\
\hline 
& & Bi & Fe & O & Bi & Fe & O \\
\hline 
cubic & $Z^*_{j}$ & --- & --- & --- & $6.32$ & $4.55$ & $-3.06$ \\ 
& $\Delta P$ & \multicolumn{3}{c|}{---} &
\multicolumn{3}{c}{123.1~$\mu$C/cm$^2$} \\
\hline
$R\bar{3}c$ & $Z^*_{j}$ & --- & --- & --- & $4.92$ & $4.25$ & $-3.06$
\\
& $\Delta P$ & \multicolumn{3}{c|}{---} &
\multicolumn{3}{c}{101.2~$\mu$C/cm$^2$} \\
\hline
$R3c$ & $Z^*_{j}$ & $4.28$ & $3.26$ & $-2.50$ & $4.37$ & $3.49$ & $-2.61$ \\
& $\Delta P$ & \multicolumn{3}{c|}{84.2~$\mu$C/cm$^2$} &
\multicolumn{3}{c}{87.3~$\mu$C/cm$^2$} \\
\hline\hline
\end{tabular}
\end{table}

Table~\ref{table:BECs} summarizes our calculated BECs for the three
different structures studied in this work. The BECs are calculated by
finite differences: the ions are displaced slightly along
[111] from their equilibrium positions, and the resulting change in
polarization is calculated using the Berry-phase method described in
Sec.~\ref{sec:method}. The displacements are chosen to be small enough
to ensure the validity of the linear treatment in
Eq.~(\ref{eqn:polchange}). (Typical displacements used here are of the
order of 0.01~\AA.) For the LSDA+U calculations we use
$U_\text{eff}$=2~eV, since this value gives an insulating solution for
all systems. (Higher values of $U_\text{eff}$ would result only in
small quantitative changes).

The highly anomalous values of the BECs computed in the cubic
structure, and the reduction of these values after freezing in the
structural distortions, are in ageement with former observations in
other ferroelectric materials.\cite{Ghosez_etc,veithen} In particular,
the large values of the Bi BECs emphasize the important role of this
ion as driving force of the ferroelectric
distortion.\cite{Seshadri_Hill} This is in some contrast to the Li ion
in the related compound LiNbO$_3$,\cite{veithen} whose effective
charge is found not to deviate appreciably from its nominal value.

In Table~IV we also report the polarization difference 
between the cubic (or $R\bar{3}c$) centrosymmetric structures 
and $R3c$. Using Eq.~(\ref{eqn:polchange}) and the $R3c$ LSDA BECs, 84.2~$\mu
C/cm^2$ is obtained along [111],\footnote{The inversion symmetry-breaking
displacements along [111] from cubic perovskite to $R3c$ are the same
as those from $R\bar{3}c$ to $R3c$.} a value consistent with the large
displacements and BECs. Yet since Eq.~(\ref{eqn:polchange}) is
only valid for small displacements and because the BECs are found to
change considerably along the path (compare the LSDA+U results for cubic, $R\bar{3}c$, 
and $R3c$ in Table~\ref{table:BECs}), this value is only a rough estimate of 
the polarization in BiFeO$_3$.  This
can be clearly seen from the larger values of 123.1 and 101.2
$\mu$C/cm$^2$ for $\Delta P$, also given in Table~\ref{table:BECs},
obtained using the cubic perovskite and $R\bar{3}c$ BECs,
respectively. The accuracy could be improved by breaking the path into
shorter segments and recomputing the BECs at intermediate points on
the path. However, it is more efficient to compute the polarization
directly from the Berry-phase theory, as will be described in the next
section.

\subsubsection{Modern theory of polarization}

An accurate quantitative method for computing the polarization to all
orders in displacement is the so-called Berry phase (or ``modern'')
theory of polarization,\cite{ksv,vks,resta} discussed in
Sec.~\ref{sec:method}. In this approach, the electronic contribution
to the polarization is calculated as a geometric phase, formally
equivalent to the sum of Wannier centers of the occupied bands (where
each center is assigned a charge $e$).\cite{ksv,vks} Because of the
Born-von Karman periodic boundary conditions employed here, there is
an ambiguity in the choice of unit cell and the total polarization may
only be determined up to an integer multiple of the polarization
quantum $e${\bf R}/$\Omega$, where $e$ is the charge of the electron,
{\bf R} is a lattice vector in the direction of
polarization, and $\Omega$ is the volume of the unit
cell.\cite{ksv,vks,resta} Thus for a given structure, the theory
yields a {\it lattice} of values. The difference in polarization between
two structures (e.g., a polar structure and its enantiomorph)
is therefore determined only up to an integer multiple of
the polarization quantum. To predict which of the values would be
obtained in a Sawyer-Tower or CV experiment, it is also necessary to
specify a switching path along which the system stays insulating in
all intermediate structural states.\cite{umesh}

In the case of BiFeO$_3$ the polarization lattice of the
$R3c$ ground state is computed to be
$(6.6+n\cdot184.2)~\mu$C/cm$^2$ along [111] within the LSDA, where $n$ is an
integer and $|e{\bf R}/\Omega|$=184.2~$\mu$C/cm$^2$ is the
polarization quantum.  Similar results are obtained with LSDA+U (see
Table \ref{table:PofU}), though a slight shift in polarization with
increasing $U_\text{eff}$ is observed. If the point $n$=0 is selected,
near-perfect agreement is obtained with the measurement of
6.1~$\mu$C/cm$^2$ by Teague {\it et al.}\cite{teague} over thirty
years ago. However, it is initially puzzling that none of the allowed
values are at all close to the estimate of
84.2~$\mu$C/cm$^2$ made in the previous section, using
Eq.~(\ref{eqn:polchange}) and based on paths involving nearby
centrosymmetric structures.

%
\begin{table}
\caption{Minimum value of the polarization lattice corresponding to
  the $R3c$ structure, computed with different values of
  $U_\text{eff}$.  All values are given modulo the polarization
  quantum.}
\begin{tabular}{l|ccc}
\hline\hline
\hspace*{2cm} & \hspace*{.7cm} & \hspace*{.7cm} & \hspace*{.7cm} \\
$U_\text{eff}$ [eV] & 0 & 2 & 4 \\
\hline
$P$ [$\mu$C/cm$^2$] & 6.6 & 2.6 & 1.2\\
\hline
\hline
\end{tabular}
\label{table:PofU}
\end{table}

This discrepancy is immediately resolved by computing the allowed values of the
polarization {\it difference} between a centrosymmetric structure
and $R3c$.  In this case it turns out that the centrosymmetric
$R\bar{3}c$ and cubic perovskite structures have a 
polarization not of zero (modulo the polarization quantum) but of half
of a polarization quantum $(92.1+n\cdot184.2)~\mu$C/cm$^2$ along 
[111].\footnote{Here we assume the volume of the centrosymmetric 
structure is equal to that of $R3c$. In general this will not be
the case (see Table~\ref{table:rbar3c} for $R\bar{3}c$), and the polarization quantum
will differ for the two structures. When computing the spontaneous
polarization, this discrepancy can be avoided by taking differences
between two endpoint phases, as will be done later in this section.}
Thus one of the allowed values for the polarization difference between
$R3c$ and the centrosymmetric reference structures, computed with the
Berry-phase theory, is 94.7~$\mu$C/cm$^2$ ($U_\text{eff}=2$~eV),
agreeing well with the linear estimate of the previous section.  That the centrosymmetric
structure has a non-zero value of polarization may at first seem
counterintuitive. However, it is readily understood using the Wannier
function reformulation of the Berry-phase expression.\cite{vks}
Making the conservative assumption that the Wannier
functions are centered on atoms with multiplicity consistent with
formal valences of $+$3 for Bi and Fe and $-$2 for oxygen; 
taking the origin at a Bi atom; and choosing as
the additional basis atoms the Fe in the center of the ideal cubic
perovskite unit cell and the three oxygens at the centers of the
faces, the Wannier sum yields a polarization of 
92.1~$\mu$C/cm$^2$ along [111] for the doubled unit cell (using
the $R3c$ volume). This value is exactly half of
a polarization quantum, the only non-zero value allowed by symmetry
for a centrosymmetric structure, \cite{vks} and also exactly the same
value that is obtained by the exact calculation of the polarization
for both centrosymmetric structures (cubic perovskite and
$R\bar{3}c$). Alternate choices of origin and basis could result in
other points on the polarization lattice, all differing by a
polarization quantum, but would never produce zero for these BECs.

\begin{figure}
\includegraphics*[width=0.95\columnwidth]{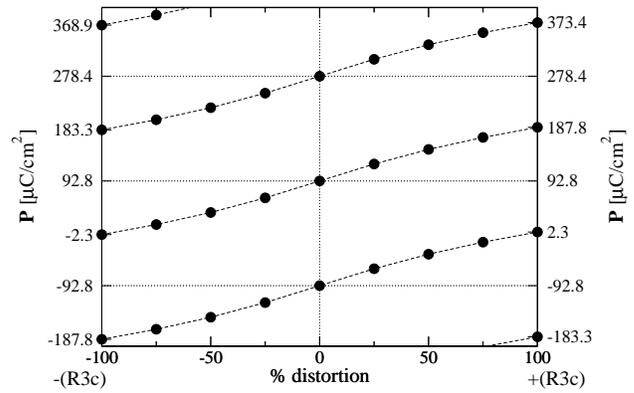}
\caption{Change in polarization ${\bf P}$ along a path from the
  original $R3c$ structure through the centrosymmetric cubic structure
  to the inverted $-(R3c)$ structure calculated with the LSDA+U method
  and $U_\text{eff}=2$~eV. The possible values of ${\bf P}$ for fixed
  distortion differ by multiples of the polarization quantum, here
  $185.6~\mu$C/cm$^2$ for $\alpha$=60$^\circ$.}
\label{fig:path}
\end{figure}

To connect the allowed polarization difference of
94.7~$\mu$C/cm$^2$ to a specific path in structural space,
we compute the polarization from the Berry-phase theory for the
endpoints and several intermediate structures along an idealized
``switching path'' connecting the positively-oriented $+(R3c)$ with
its enantiomorphic counterpart, the negatively-oriented $-(R3c)$
structure with opposite polarization, through the centrosymmetric
cubic perovskite structure. Although the actual displacements
associated with switching will obviously be more complex, the atomic
positions are assumed to travel smoothly from positive to negative
orientation through the cubic arrangement.  Also, for simplicity these
calculations assume a rhombohedral angle $\alpha$=60$^\circ$ for
$R3c$, which slightly shifts the polarization quantum to
$185.5~\mu$C/cm$^2$ and the Berry phase polarizations by a few
percent.  Since the cubic perovskite structure is metallic within
LSDA, we use the results obtained for $U_\text{eff}=2$~eV. The results
are shown in Fig.~\ref{fig:path}. For the specific continuous
insulating path connecting the $-(R3c)$ lattice point corresponding to
$-2.3~\mu$C/cm$^2$ with the $+(R3c)$ point of $187.8~\mu$C/cm$^2$, 
the polarization is seen to evolve smoothly (but
nonlinearly) through the centrosymmetric cubic structure, which has a
nonzero polarization. Subtracting the two endpoint values gives a
polarization change of $190.1~\mu$C/cm$^2$
and a predicted spontaneous polarization of half
that value, or $95.05~\mu$C/cm$^2$. (The slight difference from the value
of $94.7~\mu$C/cm$^2$ reported above is due to the assumption  $\alpha$=60$^\circ$.)
This value is consistent with measurements on
(111) oriented thin films,\cite{RameshPC} as well as
our estimate with Eq.~(\ref{eqn:polchange}).

\section{Discussion: Possibility of multiple polarization paths}
\label{subsec:connect}

In the previous section we observed that if the system switches along
the path indicated in Fig.~\ref{fig:path} (or along one that can be
continously deformed into it) the modern theory of
polarization predicts a measured spontaneous polarization of
$\sim$90-100~$\mu$C/cm$^2$, depending on the choice of
$U_\text{eff}$. In principle however, an infinite set of
polarization differences between the $+(R3c)$ and $-(R3c)$ structures
is possible, provided that a suitable pathway can be found to connect
any two endpoints.  For example, if a switching path could be found
taking $+(R3c)$ to $-(R3c)$ through an intermediate structure with
{\it zero} polarization, the measured polarization would be
$2.3~\mu$C/cm$^2$ (for $U_\text{eff}=2$~eV), consistent with
reports for bulk samples.\cite{teague} Thus although the
possibility of the small experimental values resulting from poor
sample quality cannot be ruled out, it is nonetheless interesting that
different reported values for polarization (both lower, $\approx
6~\mu$C/cm$^2$, and higher, $>$~$150~\mu$C/cm$^2$) may be explained by
different {\it switching paths}.

Specific paths connecting the smallest $+(R3c)$ polarization value
with the smallest $-(R3c)$ value have yet to be determined. One
approach to finding such a path would be to identify a centrosymmetric
reference structure with zero polarization (modulo a quantum), the path then being the
atomic displacements from $R3c$ to this structure. To get a zero
(or integer-quantum) polarization from only a small distortion of the perovskite structure would
require a somewhat drastic rearrangement of the centers of the Wannier functions. In
the present case this may be facilitated by the known multiple
valences possible for Bi and Fe. For example if the Bi ions were to
acquire an average formal charge of +4 (likely if a disproportionation
to Bi$^{3+}$ and Bi$^{5+}$ occured, leaving the Fe with a +2 charge),
the polarization of the cubic phase would be zero. 
Further, by combining a forward switching path between $R3c$
enantiomorphs through an integer-polarization  structure with another back
through a half-quantum polarization structure (e.g., $R\bar{3}c$), the
BiFeO$_3$ crystal could be taken to itself with a net transport of
electrons across the system. This suggests the intriguing possibility of
an insulating crystal with nonzero electronic conductivity.

In constructing paths corresponding to different polarizations, we do
not wish to claim knowledge of the actual experimental switching
mechanism, which is certainly much more complicated. In reality,
switching is thought to occur via domain wall motion, where the key
physics is associated with polarization reversal under an applied
electric field at interfaces between positively- and
negatively-oriented domain walls.\cite{domainwall}

Finally, an alternative explanation for the
different polarizations reported in various film and bulk samples
would be that different crystal structures are epitaxially stabilized
in the films which in turn possess significantly different
polarizations. An alternative tetragonal structure that satisfied this
requirement was discussed in a previous paper.\cite{wang} To clarify
these issues further experimental work will be necessary to determine
the sensitivity of the structure of thin film BiFeO$_3$ to substrate
and growth orientation.

\section{\label{sec:conc}Conclusions}

In summary, BiFeO$_3$ is a material of unusual interest both as a
potentially useful multiferroic, and with respect to its fundamental
polarization behavior.  A wide range of measured polarization values
have been reported, all of which are apparently permitted within the
modern theory of polarization\cite{ksv,vks,resta} by the lattice
character of the polarization. Since some of the observed values of
polarization can only be explained by switching through structures in
which the ions change their valence states, such
behavior, if experimentally verified, might be unique to
multiferroics, in which the magnetic transition metals are able to
adopt multiple values for $d$ orbital occupancy.

\begin{acknowledgments}
We thank M. H. Cohen, R. Ramesh, D. G. Schlom, and D. Vanderbilt for
useful discussions. This work is supported by the National Science
Foundation through Grants No. DMR-99-81193, NSF-MRSEC DMR-00-80008
(KMR), DMR-03-12407 (NAS) and NSF-MRSEC DMR-00-80034 (CE), and by the
Office of Naval Research through N00014-00-1-0261 (KMR).
\end{acknowledgments}


\bibliography{paper.bib}

\end{document}